# Improving the Delivery Rate of Digital Inclusion Applications for Amazon Riverside Communities by Using an Integrated Bluetooth DTN Architecture

**Ronedo Ferreira**[†], **Waldir Moreira**[††], **Paulo Mendes**[††], **Mario Gerla**[†††], and **Eduardo Cerqueira**[†]

[†]ITEC/University of Pará, Belém, Brazil,
[††]COPELABS/University Lusófona, Lisbon, Portugal
[†††]CSD/University of California, Los Angeles, USA

**Summary**
Despite the evolution in deployed infrastructure and in the way that people access information, still there are those who are socially excluded and have no access to information due to their geographic location (e.g., riverside/countryside communities). This paper proposes an extension to a DTN architecture implementation to allow the dissemination of information in such communities, including educational short-video clips and audio books. The IBR-DTN architecture is complemented with a Bluetooth Convergence Layer, to facilitate the exchange of information over this short-range wireless technology, and with a Bundle Compression mechanism that aims at improving data exchange in short-lived opportunistic contacts happening among nodes. Experiments in a small-scale testbed and in a large-scale simulator environment show that nodes are indeed able to efficiently use contact opportunities to exchange an increased amount of data, allowing people in riverside communities to receive more content related to digital inclusion services.
*Keywords:*
*Convergence Layers, DTN, Bluetooth, Bundle Compression and Infrastructureless, Video, Audio, Text Messages and Riverside People.*

## 1. Introduction

The evolution of wireless technologies increased the popularity of mobile devices and allowed people to have access to different types of applications, such as video-based education and e-health. Despite this evolution, in isolated rural areas the lack of Internet connectivity and/or network infrastructure contributes to the social exclusion of many people. Therefore, Delay/Disruption-Tolerant Networking (DTN) is a suitable solution to allow content distribution even in geographically isolated areas [1]. This networking paradigm can be used to transmit information in video format to illiterate people located in riverside/countryside communities as a way to facilitate their inclusion in the digital world. Information regarding disease prevention and treatment, and educational messages can be easily disseminated in these areas, helping improving the life quality of these people.

By employing the store-carry-and-forward approach, DTNs can efficiently deliver non-real time data (e.g., educational short videos and heath advertisement) through hop-by-hop opportunistic connections [2]. However, current DTN architecture implementations, such as DTN2 [3], Bytewalla [4], and IBR-DTN [5], are only suitable for environments presenting a certain level of wireless infrastructure, including Wi-Fi Access Points (APs) and routers. Needless to say, even the minimum level of infrastructure cannot be found in many rural scenarios, where, for instance, people only have mobile devices with Wi-Fi, Bluetooth, and GSM support (not APs at home or school).

Bluetooth technology is already present in most current smartphones and may be used as a powerful solution to provide DTN communication in isolated rural regions. The concept behind Bluetooth is to support universal short-range wireless capabilities using the 2.4GHz globally unlicensed low power band. The IBR-DTN architecture is a strong candidate to allow information exchange in such infrastructureless rural areas. Nonetheless, the architecture does not support Bluetooth connections, which limits its use in the aforementioned riverside/countryside communities. In this case, it is necessary to support Bluetooth by creating a Bluetooth Convergence Layer (BCL), thus allowing the direct exchange of information between DTN nodes in infrastructureless areas. This node-to-node communication allows the direct exchange of bundles given the implementation of the Bundle Layer [4]. Additionally, it is important to develop a compression mechanism for DTN connections in order to allow the transmission of more data per instant time.

This paper extends the IBR-DTN implementation by including BCL and Bundle Compression modules to allow node-to-node communication in riverside communities with improved data transmission during a contact opportunity. The impact and benefit of the proposed solution is implemented and evaluated in a small-scale testbed and in a large-scale simulator environment. The results show that the proposed solution increases the amount of data received during a contact over the Bluetooth technology.

This paper is structured as follows. The related work is presented in Section 2. Section 3 details the implementation of the Bluetooth Convergence Layer Modules. Results are discussed in Section 4. Finally, Section 5 presents the conclusions and future works.

## 2. Related Work

As mentioned before, there are different implementations of the DTN architecture, each with its own particularities and application scenarios. This section describes the three main state-of-the-art DTN architecture implementations available, namely, DTN2 [1], Bytewalla [2], and IBR-DTN [3].

DTN2 was developed in the context of the DTN research group. It was mainly implemented as a testing platform for basic DTN functionalities. It follows the standard and incorporates components of the DTN architecture. Still, it does not provide support for node-to-node communication using Bluetooth, and does not have a compression module for DTN bundles to become weightless. These features would consequently extend the capabilities of this DTN architecture implementation to be employed in isolated rural areas.

The Bytewalla project was developed to provide connectivity to rural villages by means of Android phones carrying an implementation of the DTN Bundle protocol. In general terms, the lack of a link between a village without Internet access and a city with Internet connectivity is replaced by a data mule, i.e., a person carrying the Android phone between both ends. When the mule is in the village, it receives a set of bundles from the Server via an 802.11 AP, and when it arrives to the city, it will connect to the gateway APs to upload the received data. However, Bytewalla does not support node-to-node communication using Bluetooth as a way to allow content to be disseminated in riverside communities. Additionally, it does not implement an API or a framework that allows working with Bluetooth (in the context of Bytewalla, this technology is only used for neighbor discovery). Finally, this DTN architecture implementation requires the Bundle Compression module to make the most out of the contacts happening between the devices carried by the people acting in these communities.

IBR-DTN was initially designed for embedded devices such as routers, and was later extended to the Android mobile platform. As with the aforementioned DTN architecture implementations, IBR-DTN only allows the exchange of information through APs and its Wi-Fi module, even though it has a modular architecture with features that could take advantage of node-to-node communication using Bluetooth. The IBR-DTN should be extended to allow Bluetooth communication and provide a mechanism to compress the data during information exchange.

The above implementations of the DTN architectures have limitations in the scenario we target: riverside/countryside communities. They rely on some level of infrastructure networking to allow data exchange among nodes, which is not a reality in such locations. Additionally, since they overlook the use of Bluetooth or do not provide support to such technology for the Android platform, this also hinders the deployment of these DTN architecture implementations in the target communities.

Among these implementations, IBR-DTN stands out, as it has been developed in modules. This allows for the implementation to be easily extended for our purposes: i) the definition of a BCL module to provide node-to-node communication using Bluetooth, and ii) implementation of a bundle compression scheme to make the most use out of any contact opportunity.

## 3. Proposed Solution

This section describes the BCL module, which is incorporated into the IBR-DTN architecture, and how its major modules interact with each other. This section also presents a brief description of the BCL components, including Bluetooth elements and the Bundle compression scheme. Figure 1 shows the IBR-DTN components together with BCL, where more details about IBR-DTN modules, interfaces, and parameters can be found [3]. All events in the IBR-DTN architecture are managed by the EventSwitch module, including those between BCL and other components.

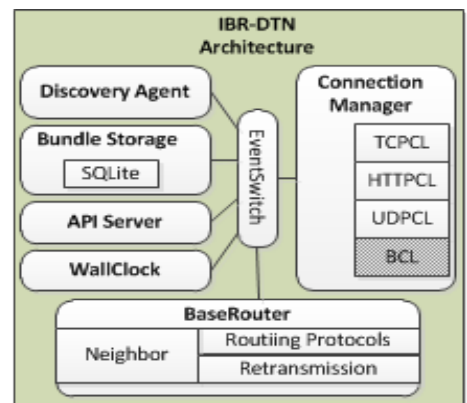

Fig. 2 BCL components

The IBR-DTN Connection Manager is responsible for creating configuration interfaces to allow external protocols to exchange DTN-based data. In the context of BCL, the IBR-DTN Connection Manager is used to configure key communication parameters, including TTL, Discovery Bluetooth Time, and DTN endpoint identifiers (EIDs). After the configuration process, BCL can transfer Bundles to neighboring nodes using Bluetooth.

The following subsections present our proposed BCL along with its components.

## 3.1 Bluetooth Convergence Layer (BCL) Components and Interactions

The BCL module extends IBR-DTN with Bluetooth communication and Bundle/data compression schemes. Thus, it will be possible to provide data exchange in DTN-based rural areas using a Bluetooth infrastructureless wireless approach. Figure 2 shows the three main BCL components, namely, Bluetooth Connection Discovery, Bluetooth Bundle Adaptation, and Bluetooth Compression Control.

The Bluetooth Connection Discovery component is responsible for discovering DTN participant nodes in the Bluetooth transmission range. The Bluetooth Bundle Adaptation adjusts all Bundles to be disseminated using Bluetooth connections, including the management of Bluetooth addresses. Finally, the Bluetooth Compression Control compresses all Bundles (stored in the Storage System) before their transmissions.

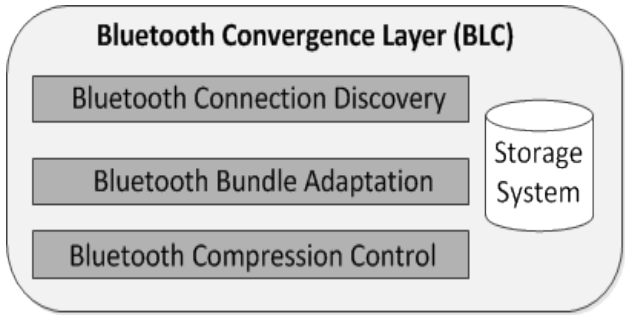

Fig. 2 BCL components

The UML 2.0 activity diagram in Figure 3 shows how a digital inclusion application (e.g., educational video clips or text books) can be delivered in the proposed IBR-DTN architecture with BCL support.

After detecting all neighboring DTN peers, the BCL Bluetooth Connection Discovery triggers the IBR-DTN Discovery Agent and informs it of all Bluetooth peers in the short-range wireless environment. This procedure is refreshed/repeated every 10 seconds in order to keep the system updated about all DTN nodes in a certain area. This parameter can be adjusted according to different scenarios or applications. If there are any Bundles (stored and managed by the Storage System) to be delivered to a (set of) DTN peer(s), the routing protocol is triggered to start the route selection.

The IBR-DTN Base Router component can be configured to control different routing protocols. This article only considers Epidemic [5] routing, as it is suitable for digital inclusion services in riverside communities where content must be delivered to as many subjects as possible. Therefore, it receives events about arriving or departing DTN nodes from the IBR-DTN Discovery Agent and notifies the routing protocol when new Bundles arrive in the Storage System. If there is no Bundle to be transmitted, the connection is closed.

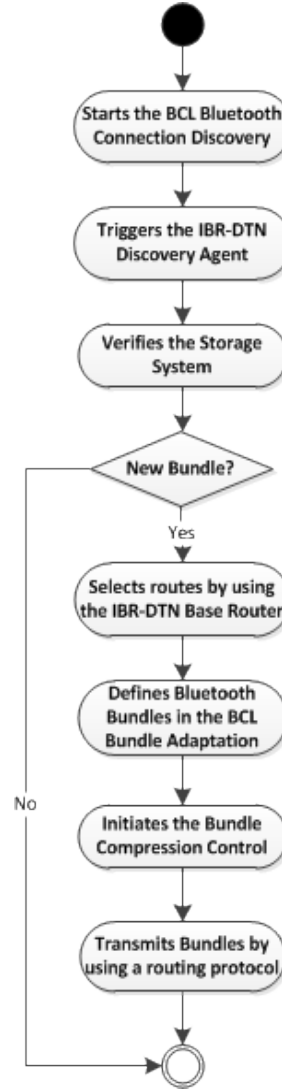

Fig. 3 Bundle Transmission in an integrated IBR-DTN BCL architecture

The Storage System is responsible for buffering and managing Bundles for an extended period of time, following the store-carry-and-forward paradigm. This component is implemented using SQLite because it is a well-known lightweight Android database that permits the storage and replacement of Bundles in a smartphone's persistent storage system, such as a Hard Disk (HD) or Secure Digital (SD) card. As presented before, after receiving new Bundles the Storage system triggers an event, which is forwarded by IBR-DTN to the Base Router to define available routes.

In contrast to current DTN architecture implementations, BCL Bluetooth Bundle Adaptation organizes the data to be transferred in Bundles (including the management of addresses) and interacts with the Epidemic routing protocol to deliver the Bundles using Bluetooth interfaces, not Wi-Fi as in current DTN solutions. The BCL Bluetooth Bundle Adaptation could be implemented together with any other protocols, such as dLife [6] as done in the SocialDTN implementation [7]. After that, the Bundles are forwarded to the BCL Bluetooth Compression Control module to compress them (including payloads and primary blocks) before their transmissions.

The BCL Bluetooth Compression Control is responsible for compressing all Bundles in order to allow node-to-node communication in riverside communities with improved data transmission during a contact opportunity (i.e., reducing the number of bytes used in each Bundle block). This functionality is very important in scenarios with short contact periods and with low transmission rate technologies, such as Bluetooth. This component uses an algorithm for lossless storage: the Deflate algorithm is used in this article because it can achieve a good performance by identifying and eliminating statistical redundancy [8].

A degree of block compression is provided by the design of the payload and primary blocks. The former contains the application data carried by the Bundle, which is compressed using the Deflate algorithm. The primary block contains the scheme names and scheme specification parts of the four EIDs. Up to eight EIDs are concatenated at the end of the block in a variable length character array, called a dictionary, enabling each EID to be represented by a pair of integers indicating the offsets within the dictionary of the EID's scheme name and scheme specification part.

When the total length of the dictionary is less than 127 bytes, all eight offsets can be encoded into just eight bytes. However, these strategies do not prevent the scheme names and, especially, the scheme specification parts themselves from being lengthy strings of ASCII text. Therefore, it is still possible for the length of a bundle's primary header to be a very large fraction of the total length of the bundle when the bundle's payload is relatively small.

Finally, after accomplishing the compression schemes, the BCL Bluetooth Compression Control triggers the IBR-DTN Base Router to initiate the routing protocol and the data delivery process to DTN peers by using Bluetooth.

## 4. Performance Evaluation

This section presents the performance evaluation results of the proposed integrated IBR-DTN BCL architecture. The objective is to measure the impact and the benefit of using Bluetooth to deliver data with and without a compression scheme for DTN communities. A testbed was implemented to collect results from a real environment, as well as to calibrate the simulator with real measurement data. In addition, a simulation environment was configured to analyze the performance of the proposed solution in large-scale scenarios. Examples of educational video-based flows that can be used for digital inclusion services are available in the Khan Academy[1].

### 4.1 Testbed

A testbed was setup at the Federal University of Para (UFPA – Brazil) campus to understand the impact of the proposed solution in a real network as presented in Figure 4. The IBR-DTN BCL architecture was implemented in six Samsung WI8 150 smartphones with Android 4.0; their characteristics are presented in Table 1.

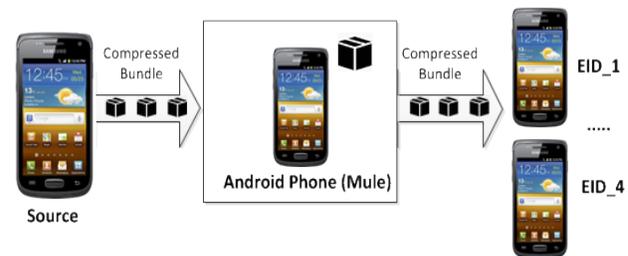

Fig. 4 Testbed scenario

Table 1: Smartphones Samsung WI8150

| Parameters | Values |
|---|---|
| OS | Android OS v4.0 |
| CPU | 1.4 GHz Scorpion |
| Battery | Li-Ion 1500 mAh battery |
| Memory | 1 GB storage, 512 MB RAM, 2 GB ROM |
| Bluetooth | v3.0 with A2DP |

---

[1] www.khanacademy.org

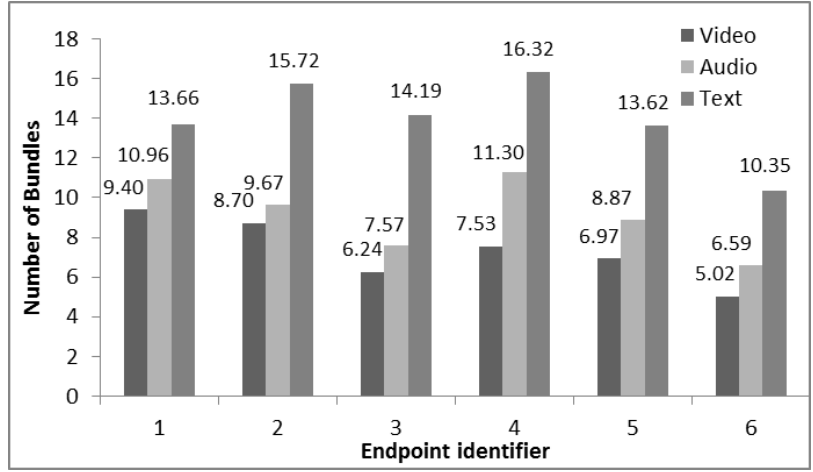

Fig. 5 Number of Compressed Bundle Received for Each Mobile Device/EID

The Epidemic routing protocol was configured to select peers and transmit Bundles using a Bluetooth 3.0 interface. Table 2 describes the testbed parameters used during the experiments, including information about buffer size (assuming that 1000MB is a buffer size large enough to accommodate many digital inclusion video flows with an average size of 5MB) and compression levels (from 5% to 25%). During the experiments, users carried the smartphones and walked about randomly in a 40-square meter office the size of some piers, schools, or houses in the Amazon.

Table 2: Testbed Parameters

| Parameters | Values |
|---|---|
| Number of Nodes | 6 |
| Buffer Size | 1000MB |
| Node Speed | 0.8m/s - 1.4m/s |
| Bundle Size | 100Kb-5000KB |
| Contact Time | 1s - 600s |
| TTL | 3038s |
| Transmission range | 10 m |
| Transmission rate | 256Kbps |
| Compression levels | 5%-25% |
| Discovery Transmission Refresh Interval | 10s |

The results show that the proposed compression scheme is important for delivery of more Bundles during a Bluetooth contact. The time required for transferring one Bundle of a video-clip when the IBR-DTN BCL is configured with a compression algorithm is 7s. Another important performance evaluation result is that IBR-DTN BCL requires only 3s for delivering 10 Bundles of texts.

4.2 Simulation Environment

Large-scale simulations were carried out to analyze the benefits of the proposed solution using Simulator ONE [9]. The bundle delivery rate and Bluetooth channel characteristics were collected in the testbed and used to configure the simulation environment.

Figure 7 shows possible DTN scenarios that can take place in the riverside communities of the Amazon. Such scenarios are defined based on the behavior, interaction, and contact of the nodes, as well as the transport, mobility, and schedule models of the boats. The scenario is composed of two main piers with Wi-Fi DTN servers that are responsible for transmitting data (text, audio, and videos) to specific EIDs.

The network communication in the remaining piers, boats, and communities is performed through Bluetooth. Three different digital applications are defined for the experiments as follows: text (Bundle size of 100Kb – e.g., email, HTTP), audio (Bundle size of 3000Kb – e.g., audio books), and video (Bundle size of 5000Kb – e.g., e-learn videos).

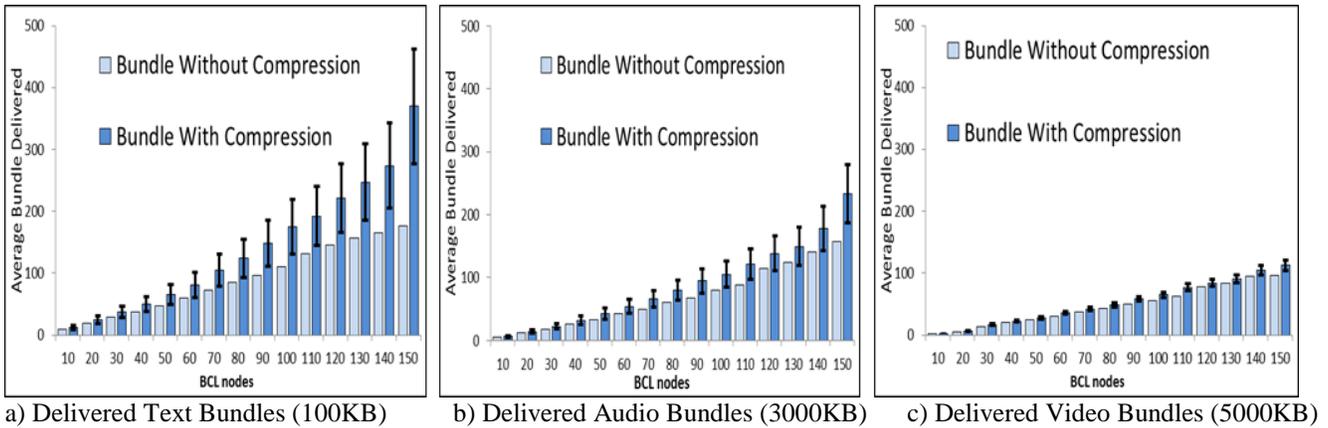

a) Delivered Text Bundles (100KB)    b) Delivered Audio Bundles (3000KB)    c) Delivered Video Bundles (5000KB)

Fig. 6 – Average Bundle Delivered for each application

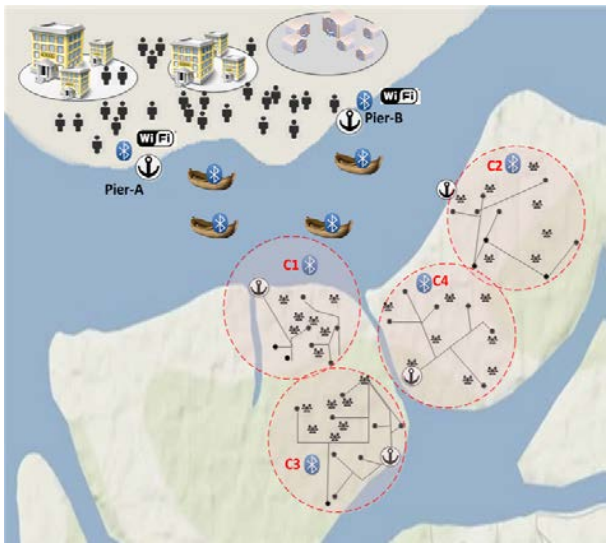

Fig. 7 Simulated scenario

The simulator was configured to run 100 simulations during 10 days and the results are presented with a confidence interval of 95%. For each simulation, the number of nodes receiving the content from the Wi-Fi DTN server in Pier A and Pier B and exchanging Bluetooth contacts before moving, by boat, to their communities varies from 10 to 150. For instance, in some experiments, Pier A and Pier B support up to 75 people before the boarding time (total of 150 nodes). The number of subjects/nodes that each boat can transport to each one of the four riverside communities at a specific time ranges from 1 to 10.

Four riverside communities are specified in the environment, namely, C1 to C4 as presented in Figure 6. Each DTN peer is configured with the Working Day Movement model and his/her speed varies from 0.8 and 1.4 m/s. Moreover, 10 boats are used in the experiments.

The boats travel following the Shortest Path Map Based Movement approach, where they randomly choose a place and use the shortest path to reach it. Their speed ranges from 5 to 7 m/s and they can wait for passengers in each pier (i.e., boarding time) from 1200 to 1800s. Table 3 summarizes the simulator parameters.

Table 3: Simulation Parameters

| Parameters | Values |
|---|---|
| Number of Nodes | 10-150 |
| Buffer Size | 1000MB |
| Node Mobililty | 0.8m/s – 1.4m/s |
| Boat Mobility | 5m/s – 7m/s |
| Bundle Text Size | 100Kb |
| Bundle Audio Size 128Kb/s ~ 3 min | 3000Kb |
| Bundle Video Size 3GP - ~ 5 min | 5000Kb |
| Contact Time | 1s - 600s |
| TTL | 24h |
| Transmission range | 10 m |
| Transmission rate | 256Kbps |

As presented in Figure 6, the results show that, on average, when the Bundle Compression scheme is used for delivering text, audio, and video, it reduces the Bundle sizes by 50%, 7%, and 5%, respectively. Thus, the proposed BCL implementation transmits more data during a Bluetooth contact. For instance, for audio files, it is possible to deliver 10% and 21% more Bundles when the compression model is configured in the system with 70 and 110 nodes. The best results are achieved from the distribution of text files in a scenario with 150 nodes, where 50% more bundles are delivered.

The results also show that it is possible to receive 14% more video Bundles when the compression approach is used in an environment with 150 peers. Since the video files already have a high compression level (MEncoder – 3GP), the performance of the Deflate algorithm is reduced as happened when transmitting video flows in a scenario with only 10 nodes.

The use of the compression model together with a Bluetooth IBR-DTN is an important solution for transmission of more data in dynamic and short-contact DTN Bluetooth scenarios. New compression algorithms could be used to reduce Bundle files and improve the bundle delivery rate.

## 5. Conclusion and Future Work

The use of DTN is a suitable approach to providing digital inclusion services in areas with intermittent (or inexistent) Internet connectivity. The implementation of a Bluetooth and wireless infrastructureless scheme increases the possibility of delivering content to riverside or countryside communities. This article presented an extended version of the IBR-DTN architecture with Bluetooth support. Additionally, it integrated a Bundle Compression Control scheme into the BCL approach to improve the number of Bundles received during short-lived contacts. Performance evaluation results show that the use of a compression algorithm, such as Deflate, can increase the number of Bundles delivered for text, audio, and video applications by 55%, 7%, and 5%, respectively, when the scenario is composed of 150 nodes.

For future work, experiments with a large-scale testbed with different applications will be performed. Different compression algorithms and network coding techniques will also be evaluated. Finally, the performance of DTN systems is associated with their routing protocols; therefore, IBR-DTN BCL will be analyzed together with different routing schemes.

**Acknowledgments**
Thanks are due to the National Council for Scientific and Technological Development (CNPq), to the Amazon Research Foundation (FAPESPA), to Lemann Foundadtion, and to FCT for financial support of the User-Centric Routing project (PTDC/EEA- TEL/103637/2008) and Waldir Moreira's PhD grant (SFRH/ BD/62761/2009).

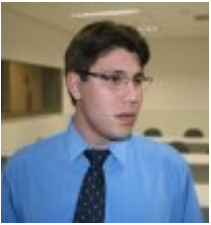
**Ronedo Ferreira** received his BSc in Computer Engineer from the University Institute of Higher Studies of the Amazon (IESAM), Belém in 2010. He is now a MSc student of the Electrical Engineer Postgraduation Program (PPGEE) of the Federal University of Pará (UFPA), Belém. His interests include DTN, Multimedia, and Performance Evaluation.

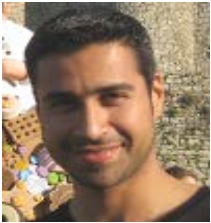
**Waldir Moreira** received his Master's degree in Computer Science from the Postgraduate Program of Computer Science (PPGCC) at Universidade Federal do Pará (UFPA), Belém/PA, Brazil. Currently, he is in the fourth year of the MAP Doctoral Programme in Telecommunications (MAP-tele) at Universidade de Aveiro (UA) and is a member of the Internet Architectures and Networking Lab (IANLab) of the Research and Development in Informatics Systems and Technologies Labs (SITILabs) at Universidade Lusófona de Humanidades e Tecnologias (ULHT).

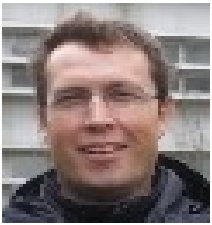
**Paulo Mendes** studied Informatics Engineering at University of Coimbra receiving a diploma degree in 1993, and at the Technical University of Lisbon receiving a M.Sc. degree in 1998 in Electrical and Computer Engineering. In 2004 he got his Ph.D. (summa cum laude) degree in Informatics Engineering from the University of Coimbra, having performed his Ph.D. thesis work as a visiting Scholar at Columbia University, New York. He is now an Associate Professor in University Lusofona. His publication record includes over 100 scientific publications, and 14 international patents. His research and teaching interests are in the area of Self-organizing Pervasive Wireless Sensing Systems, with application to mHealth and Internet of Things.

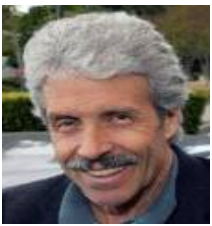
**Mario Gerla** received a graduate degree in engineering from the Politecnico di Milano, in 1966, and the M.S. and Ph.D. degrees in engineering from UCLA in 1970 and 1973, respectively. From 1973 to 1976, Dr. Gerla was a manager in Network Analysis Corporation, Glen Cove, NY, where he was involved in several computer network design projects for both government and industry, including performance analysis and topological updating of the ARPANET under a contract from DoD. From 1976 to 1977, he was with Tran Telecommunication, Los Angeles, CA, where he participated in the development of an integrated packet and circuit network. Since 1977, he has been on the Faculty of the Computer Science Department of UCLA. His research interests include the design, performance evaluation, and control of distributed computer communication systems and networks. His current research projects cover the following areas: design and performance evaluation of protocols and control schemes for Ad Hoc wireless networks; routing, congestion control and bandwidth allocation in wide area networks, and; traffic measurements and characterization.

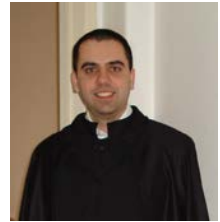
**Eduardo Cerqueira** received his PhD in Informatics Engineering from the University of Coimbra, Portugal, in 2008. He is an associate professor at the Faculty of Computer Engineering of the UFPA in Brazil. His publications include four books, four patents and over than 100 papers in national/international refereed journals/conferences. He is involved in the organization of several international conferences and workshops, including Future Multimedia Networking, Latin America Conference on Communications and Latin American Conference on Networking. He has been serving as a Guest Editor for many special issues of various peer-reviewed scholarly journals. His research involves multimedia, Future network, Quality of Experience, Mobility.